\documentclass{article}
\usepackage{graphicx}  
\usepackage{amsmath}   
\usepackage[compress]{cite}
\usepackage{amssymb}   
\usepackage{bm} 
\usepackage{dcolumn}
\usepackage{color}
\usepackage[dvipsnames]{xcolor}
\usepackage{mathrsfs}
\usepackage{gensymb}
\usepackage{float}
\usepackage{amsfonts}
\usepackage{varioref}
\usepackage{textcomp}
\usepackage{enumerate}
\usepackage{physics}
\usepackage{tikz}
\usepackage{multicol}
\usepackage{soul}
\usepackage{notoccite}
\usepackage{graphicx}
\usepackage[caption=true]{subfig}
\RequirePackage[colorlinks,citecolor=Cyan,urlcolor=Purple,linkcolor=Blue]{hyperref}
\allowdisplaybreaks
\usepackage{ccfonts,palatino}

\DeclareMathAlphabet{\mathbf} {OT1}{cmbr}{bx}{n}
\DeclareMathAlphabet{\mathbold}{OML}{cmbrm}{b}{it}
\setstcolor{red}
\labelformat{section}{Section #1}
\labelformat{subsection}{Section #1}
\labelformat{subsubsection}{Section #1}
\labelformat{subsubsubsection}{Section #1}
\labelformat{equation}{Eq.~(#1)}
\labelformat{figure}{Fig.~[#1]}
\labelformat{subfigure}{Fig.~[\thefigure#1]}
\labelformat{table}{Tab.~#1}
\labelformat{appendix}{Appendix #1}
\definecolor{lime}{HTML}{A6CE39}
\DeclareRobustCommand{\orcidicon}{
	\begin{tikzpicture}
		\draw[lime, fill=lime] (0,0)
		circle [radius=0.16]
		node[white] {{\fontfamily{qag}\selectfont \tiny ID}};
		\draw[white, fill=white] (-0.0625,0.095)
		circle [radius=0.007];
	\end{tikzpicture}
	\hspace{-2mm}
}
\foreach \x in {A, ..., Z}{\expandafter\xdef\csname orcid\x\endcsname{\noexpand\href{https://orcid.org/\csname orcidauthor\x\endcsname}
		{\noexpand\orcidicon}}
}

\title{\textsc{Shadow of Regular Black Hole in Scalar-Tensor-Vector Gravity theory}}
\author{ Subhadip Sau\footnote{\color{Blue}subhadipsau2@gmail.com}~$^{1,2}$\orcidA{} and John W. Moffat\footnote{\color{Blue}jmoffat@perimeterinstitute.ca}~$^{3,4}$\\
	{$^{1}$\small{\it Department of Physics, Jhargram Raj College, Jhargram, West Bengal-721507}}\\
	{$^{2}$\small{\it School of Physical Sciences, Indian Association for the Cultivation of Science,}} \\ {\small{\it 2A \& 2B Raja S. C. Mullick Road, Kolkata-700032,India}}\\
	{$^{3}$\small{\it Perimeter Institute for Theoretical Physics, Waterloo, Ontario N2L 2Y5, Canada}}\\
	{$^{4}$\small{\it Department of Physics and Astronomy, University of Waterloo, Waterloo, Ontario N2L 3G1, Canada}}}
\begin{document}
	
	\maketitle
	\begin{abstract}
		We investigate the shadow cast by a regular black hole in scalar-tensor-vector mOdified gravity theory. This black hole differs from a Schwarzschild-Kerr black hole by the dimensionless parameter $\beta$. The size of the shadow depends on this parameter.  Increasing the value of the parameter $\beta$ shrinks the shadow. A critical value of the parameter $\beta$ is found to be $\beta_{\rm crit}=0.40263$. The shadow for the horizonless dark compact object has been analysed for the static, spherically symmetric case and compared with M87* and Sgr A* data. Shadow observables have been determined in the context of the regular black hole and used for obtaining the energy emission rate. The peak of the energy emission rate shifts to lower frequency for the increasing value of the parameter $\beta$.
	\end{abstract}
	
	
	\newpage
	\tableofcontents
	\section{Introduction}
	One of the most remarkable predictions of the general theory of relativity is the occurrence of black holes. The recent observations by the Event Horizon Telescope (EHT) collaboration\cite{Fish:2016jil,EventHorizonTelescope:2019dse,EventHorizonTelescope:2019uob,EventHorizonTelescope:2019jan,EventHorizonTelescope:2019ths,EventHorizonTelescope:2019pgp,EventHorizonTelescope:2019ggy} and the detection of gravitational wave signals by the Laser-Interferometer Gravitational Wave-Observatory (LIGO) and Virgo \cite{LIGOScientific:2016aoc,LIGOScientific:2020stg,PhysRevLett.125.101102}, corroborate the existence of these celestial objects.  Despite its success, the theory of general relativity is not flawless. The two major drawbacks of this theory are the presence of singularities\cite{GEROCH1968526,PhysRev.187.1784} in the theory and the lack of observational data verifying the existence of the dark sector\cite{Souza18}. The research community is divided into two groups regarding this issue\cite{MARTENS2020237,NOJIRI201044,Calmet:2017voc,Sanders:2006sz,Nojiri:2008nt,Nojiri:2008ku}. Either dark matter exists or alternatively Einstein's gravitational theory has to be modified. Despite numerous attempts to find the existence of the dark sector, in particular, the dark matter, all experimental attempts to detect dark matter have until now failed\cite{baudis_2018,Liu:2017drf}. This motivates us to explore the nature of black holes in a theory where these above mentioned ambiguities are removed. One of the successful approaches towards this goal has been developed by one of the authors\cite{Moffat:2005si}. The theory is popularly known in the literature as the scalar-tensor-vector gravity (STVG) theory and MOdified gravity (MOG). The solar system observations\cite{Moffat:2014asa}, cosmological observations\cite{Davari:2021mge}, galaxy rotation curves\cite{10.1093/mnras/stt1670,Moffat:2014pia,Moffat:2013sja} and the dynamics of galaxy clusters\cite{Brownstein:2007sr,10.1093/mnras/stu855} have all been satisfactorily explained by the MOG. It has also been successful in describing structure growth, the matter power spectrum, and cosmic microwave background (CMB) acoustical and angular power spectrum data\cite{Moffat:2014bfa,Moffat:2007ju,Moffat:2011rp,Moffat:2021log}. Observational signatures and constrains of the black holes and other compact objects as appearing in MOG theory have been discussed in the  literature\cite{Hu:2022lek,Qin:2022kaf,DellaMonica:2021xcf}. To distinguish the MOG theory from general relativity, EHT observational data have been used to study the shadow cast by the supermassive MOG black holes Sgr A* and M87*\cite{Moffat:2019uxp}.

	As a result of lensing phenomena\cite{doi:10.1126/science.84.2188.506,PhysRevD.62.084003,Perlick:2010zh,Cunha:2018acu}, the black hole scatters the higher angular momentum photons from the source, sending them to the distant observer, while the photons with less angular momentum fall into the black hole and create a shadow zone and a possible light ring. The black hole shadow, which develops next to the event horizon, gives us a general notion of the fundamental geometrical structure of horizons\cite{Bronzwaer:2021lzo}. A review of these developments can be found in \cite{Perlick:2021aok}. Sagittarius A*, the supermassive black hole at the heart of our galaxy, and M87* at the galactic centre of M87 have both been confirmed by the EHT astronomical observations\cite{Fish:2016jil,EventHorizonTelescope:2019dse,EventHorizonTelescope:2019uob,EventHorizonTelescope:2019jan,EventHorizonTelescope:2019ths,EventHorizonTelescope:2019pgp,EventHorizonTelescope:2019ggy}. A two-dimensional dark disc encircled by bright rings is the black hole's observable appearance. The light rings are photon orbits, while the dark area represents the black hole shadow. The accreted matter around the black hole has an impact on how the shadow is shaped. Since the black hole's shadow carries the geometry of the surrounding region in its shape and size, it is considered a helpful tool for determining the black hole's spin and other deformation characteristics and parameters\cite{Kumar:2018ple,Ghosh:2020spb,Afrin:2021imp,Ghosh:2022kit}. This in turn can help to distinguish and test general relativity and other alternative theories\cite{Younsi:2021dxe,Psaltis:2018xkc,Mizuno:2018lxz,Stepanian:2021vvk,KumarWalia:2022aop,Vagnozzi:2022moj,Banerjee:2022iok,Banerjee:2022jog,PhysRevD.100.024028,Bambi:2019tjh,Vagnozzi:2019apd,Allahyari:2019jqz,Khodadi:2020jij,Roy:2021uye,Shaikh:2021yux,Ghosh:2022gka,Bernardo:2022acn,EventHorizonTelescope:2020qrl}. The black hole candidates display a significant rotation. Our main goal will be to explore the rotating black holes in MOG/STVG theory. To evade the problem of a singularity, we will focus our study on regular solutions in the STVG/MOG theory of gravity. One of the crucial methods for obtaining information about a black hole is the study of its shadow\cite{PhysRevD.88.064004,Atamurotov:2013dpa,Abdujabbarov:2012bn,Atamurotov:2021cgh,Papnoi:2021rvw,Lee:2021sws}. Earlier attempts have been made to analyse the shadows of regular black holes\cite{universe5070163,GHOSH2020115088,PhysRevD.100.124024,PhysRevD.93.104004,PhysRevD.97.064021,Li_2014,Banerjee:2022bxg,Banerjee:2022iok}. Many of these regular black holes arise from gravity coupled to non-linear electrodynamics. However, the electrical charge of black holes is expected to have negligible effect on the geometry of spacetime\cite{10.1093/mnras/stz1904}. In STVG/MOG theory, the regular black hole solutions are obtained from a purely gravitational theory and can be potential candidates for astrophysical black holes.

	The paper is organized as follows: In \ref{Sec2} a brief introduction to the STVG/MOG gravitational action and field equations is presented.  The static regular MOG compact object is discussed in \ref{S_MOG}. In \ref{Sec-SS}, we investigate the regular MOG spherically symmetric solution and analytically derive the critical value of the parameter $\beta$. We also derive the parameter dependence of the black hole horizon, photon sphere and shadow. \ref{Sec-RS} is dedicated to a study of the regular MOG rotating solution. In \ref{Sec-NG}, we have determined the shape and size of the black hole shadow for the regular MOG rotating solution along with the observables associated with it. Finally we have calculated the energy emission rate for the concerned black hole with the help of associated observables in \ref{Sec-EM}.
	
	Throughout the paper, we will use mostly the positive metric convention assuming the velocity of light to be unity ($c=1$).

	\section{STVG action and field Equations}\label{Sec2}
	
	The action for MOG/ STVG theory is
	\begin{flalign}
		S=S_{\rm GR}+S_{\phi}+S_{S}+S_{M}
		\label{action_1}
	\end{flalign}
	where
	\begin{flalign}
		S_{\rm GR}&=\dfrac{1}{16\pi}\int d^{4}x \sqrt{-g} \dfrac{1}{G}R\\
		S_{\phi}&=-\int d^{4}x \sqrt{-g}\left(\dfrac{1}{4}B^{\mu\nu}B_{\mu\nu}-\dfrac{1}{2}{\mu}^{2}\phi^{\mu}\phi_{\mu} - J^\mu\phi_\mu\right)\\
		S_{S}&=\int d^{4}x \sqrt{-g} \dfrac{1}{G^{3}}\left(\dfrac{1}{2}g^{\mu\nu}\nabla_{\mu}G\nabla_{\nu}G-V(G)-JG\right) +\int d^{4}x\sqrt{-g} \dfrac{1}{{\mu}^{2}G} \left(\dfrac{1}{2}g^{\mu\nu}\nabla_{\mu}{\mu}\nabla_{\nu}{\mu}-V(\mu)\right)
	\end{flalign}
	Here $g_{\mu\nu}$ is the spacetime metric, $g$ is the determinant of the metric, $R$ is the Ricci scalar, $\phi^{\mu}$ is a proca-type massive vector field such that $B_{\mu\nu}=\partial_{\mu}\phi_{\nu}-\partial_{\nu}\phi_{\mu}$, $G(x)$ and $\mu(x)$ are scalar fields and $V(G)$ and $V(\mu)$ are the corresponding potentials. $S_{M}$ is the matter action.
	The energy-momentum tensor for the gravitational source can be written as
	\begin{flalign}
		T_{\mu\nu}=T_{\mu\nu}^{M}+T_{\mu\nu}^{\phi}+T_{\mu\nu}^{S}
	\end{flalign}
	where
	\begin{subequations}
		\begin{flalign}
			T_{\mu\nu}^{M}=-\dfrac{2}{\sqrt{-g}}\dfrac{\partial S_{M}}{\delta g^{\mu\nu}}\\
			T_{\mu\nu}^{\phi}=-\dfrac{2}{\sqrt{-g}}\dfrac{\partial S_{\phi}}{\delta g^{\mu\nu}}\\
			T_{\mu\nu}^{S}=-\dfrac{2}{\sqrt{-g}}\dfrac{\partial S_{S}}{\delta g^{\mu\nu}}
		\end{flalign}
	\end{subequations}
	Here, $T_{\mu\nu}^{M}$ is the ordinary matter energy-momentum tensor, $T_{\mu\nu}^{\phi}$ is the energy-momentum tensor for the field $\phi^{\mu}$ and the scalar contribution to the energy-momentum tensor is denoted by $T_{\mu\nu}^{S}$. Moreover, $J^\mu$ and $J$ are the vector and scalar field currents, respectively.
	
	The Schwarzschild-MOG and Kerr-MOG black hole solutions can be found with the following assumptions:
	\begin{itemize}
	    \item It is assumed that the matter energy-momentum tensor $T_{\mu\nu}^{M}$ and the vector and scalar field currents $J^\mu$ and $J$ are zero.
		\item Since the effects of the vector field $\phi_\mu$ mass $\mu$ becomes prominent at kiloparsec distances from the source, the mass of the vector field is disregarded when solving the field equations for compact objects like black holes.
		\item The constant $G$ depends on the parameter $\beta=\alpha/(1+\alpha)$ by $G=G_{N}(1+\alpha)=\dfrac{G_{N}}{1-\beta}$.
		Here, $G_{N}$ is Newton's gravitational constant and we assume that $\partial_\mu G \approx 0$. The range of the dimensionless parameter $\beta$ is $0\leq \beta \leq 1$.
	\end{itemize}
	The action in \ref{action_1} assumes the following form:
	\begin{flalign}
		S=\frac{1}{16\pi G}\int \dd^{4}x \sqrt{-g}\left(R -\dfrac{1}{4}B^{\mu\nu}B_{\mu\nu}\right)
		\label{action_2}
	\end{flalign}
	Varying this action with respect to $g_{\mu\nu}$, we get the following field equations:
	\begin{flalign}
		G_{\mu\nu}=8\pi G T_{\mu\nu}^{\phi}
	\end{flalign}
	Here $G_{\mu\nu}$ is the Einstein tensor $R_{\mu\nu}-\frac{1}{2}g_{\mu\nu}R$. The energy-momentum tensor associated with vector field  $\phi_{\mu}$ is given by
	\begin{flalign}
		T_{\mu\nu}^{\phi}=\frac{1}{4\pi}\left(B_{\mu}^{~~\rho}B_{\nu\rho}-\dfrac{1}{4}g_{\mu\nu}B^{\alpha\beta}B_{\alpha\beta}\right)
	\end{flalign}
	To obtain the dynamical equation for the vector field, we need to vary the action in \ref{action_2} with respect to the vector field $\phi_{\mu}$. Such a variation leads to the following dynamical equation:
	\begin{flalign}
		\nabla_{\nu}B^{\mu\nu}=\dfrac{1}{\sqrt{-g}}\partial_{\nu}\left(\sqrt{-g}B^{\mu\nu}\right)=0
	\end{flalign}
	
	One should note here that the gravitational charge $Q_g$ associated with the MOG vector field is proportional to the mass of the gravitational source as\cite{Moffat:2014aja}
	\begin{flalign}
		Q_g=\sqrt{\alpha G_{N}}M=\sqrt{\beta(1-\beta)G_{N}} M_{\beta}
	\end{flalign}
	where $M_{\beta}=(1+\alpha)M$. The gravitational charge $Q_g$ results in the modified Newtonian acceleration for weak gravitational fields and slow particle motion:
	\begin{flalign}
	    a(r)=-\frac{G_N M}{r^2}[1+\alpha-\alpha\exp(-\mu r)(1+\mu r)]
	\end{flalign}
	For small scale objects and weak gravitational fields $\mu r << 1$ and the parameter $\alpha$ cancels, reducing the acceleration to Newtonian gravity. With
	parameter-post-Newtonian corrections this guarantees that MOG is consistent with accurate solar system experiments.

	\section{Static regular  MOG compact object}\label{S_MOG}
	
	The gravitational action for the matter-free MOG theory using non-linear field equations for the gravitational spin 1 vector field $\phi_{\mu}$ is given by \cite{Moffat:2018jmi}
	\begin{flalign}
		S_{\rm MOG}=\dfrac{1}{16\pi G}\int d^{4}x \sqrt{-g} \left[R-L(B)\right]
		\label{action}
	\end{flalign}
	where $R$ is the Ricci scalar, $L(B)$ describes the non-linear contribution of $B_{\mu\nu}=\partial_{\mu}\phi_{\nu}-\partial_{\nu}\phi_{\mu}$ with  $B=\dfrac{1}{4}B_{\mu\nu}B^{\mu\nu}$. The associated field equations are
	\begin{subequations}
	\begin{flalign}
		G_{\mu\nu}=8\pi G T^{\phi}_{\mu\nu}\\
		\nabla_{\nu}\left(\dfrac{\partial L}{\partial B}B^{\mu\nu}\right)=0\\
		\nabla_{\mu}\left( {}^{\star}B^{\mu\nu}\right)=0
	\end{flalign}
\end{subequations}
	where ${}^{\star}B^{\mu\nu}=\epsilon^{\mu\nu\rho\sigma}B_{\rho\sigma}$ is the Hodge-dual of $B^{\mu\nu}$. The energy-momentum tensor associated with the theory is given by
	\begin{flalign}
		T_{\mu\nu}^{\phi}=\dfrac{1}{4\pi}\left[\dfrac{\partial L}{\partial B}g^{\rho\sigma}B_{\mu\rho}B_{\nu\sigma} -g_{\mu\nu}L(B) \right]
	\end{flalign}
	In this theory, the gravitational constant is enhanced by $G=G_{N}(1+\alpha)$. The gravitational source charge associated with the vector field $\phi_{\mu}$ is given by
	\begin{flalign}
		Q_g=\sqrt{\alpha G_{N}} M
	\end{flalign}
	where $M$ is the mass parameter of the theory. The gravi-electric field is given by
\begin{flalign}
	E_{\rm grav}(r)= B_{01}(r)=-B_{10}(r)
	\end{flalign}
The energy-momentum tensor components are given by
\begin{flalign}
	T^{\phi 0}_{0}=T^{\phi 1}_{1}=-\dfrac{1}{4\pi}\left( E_{\rm grav}^{2}\dfrac{\partial L}{\partial B}+L(B)\right)
	\end{flalign}
To describe the non-linear system in an alternative way, one can consider the function $H$ obtained from the Legendre transformation. The function $H$ is given by
\begin{flalign}
	H=2B\dfrac{\partial L}{\partial B}-L(B)
	\end{flalign}
We assume
\begin{flalign}
	P_{\mu\nu}=\dfrac{\partial L}{\partial B} B_{\mu\nu}
	\end{flalign}
	and
	\begin{flalign}
		P=\dfrac{1}{4} P_{\mu\nu}P^{\mu\nu}=\left(\dfrac{\partial  L}{\partial B} \right)^{2} B
		\end{flalign}
	Now, $H$ can be expressed as the function of $P$. For the regular spacetime metric solution the form of the function $H(P)$ is given by
	\begin{flalign}
		H(P)=P\dfrac{\left(1-3\sqrt{-2\alpha(1+\alpha)M^{2}P} \right)}{\left(1+\sqrt{-2\alpha(1+\alpha)M^{2}P} \right)^{3}} -\dfrac{3}{2\alpha(1+\alpha)M^{2}b}\left(\dfrac{\sqrt{-2\alpha(1+\alpha)M^{2}P}}{1+\sqrt{-2\alpha(1+\alpha)M^{2}P}} \right)
		\end{flalign}
	where $b=\dfrac{\sqrt{\alpha}M}{2}$ and $P=-\dfrac{\alpha}{(1+\alpha)}\dfrac{M^{2}}{2r^{4}}$ and we have set the gravitational constant $G_N=1$. The associated Lagrangian L is provided by
	\begin{flalign}
		L(P)=P\dfrac{\left(1-8\sqrt{-2\alpha(1+\alpha)M^{2}P}-6\alpha(1+\alpha)M^{2}P \right)}{\left(1+\sqrt{-2\alpha(1+\alpha)M^{2}P}\right)^{4}}-\dfrac{3\left(-2\alpha(1+\alpha)M^{2}P \right)^{5/4}\left(3-\sqrt{-2\alpha(1+\alpha)M^{2}P} \right)}{4\alpha(1+\alpha)M^{2}b\left(1+\sqrt{-2\alpha(1+\alpha)M^{2}P} \right)^{7/2}}
		\end{flalign}

	\section{Regular MOG static spherically symmetric spacetime}\label{Sec-SS}
	
\begin{figure}[h!]
		\centering
		\includegraphics[width=8cm]{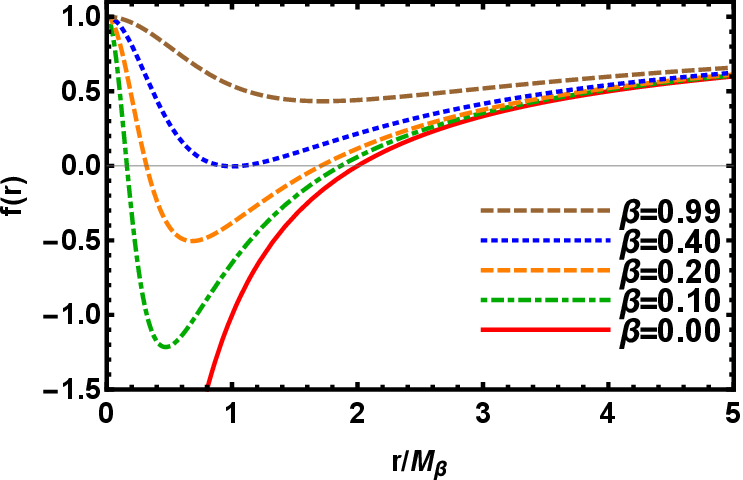}
		\caption{The zeros of the function $f(r)$ have been shown to confirm that either two horizons or no horizon are possible for the regular MOG spherically symmetric spacetime. For $\beta=0$ the spacetime becomes Schwarzschild and has a singularity at $r=0$. One horizon solution is possible for the critical value of the parameter $\beta_{\rm crit}\approx 0.40263$. It is easy to check that there is no singularity for non-zero values of parameter $\beta$.}
		\label{Zero_func}
	\end{figure}

	The MOG regular, static spherically symmetric solution can be written as\cite{Moffat:2018jmi,Ayon-Beato:1998hmi}
	
	\begin{flalign}
		ds^{2}=-f(r) dt^{2}+ \dfrac{1}{f(r)}dr^{2}+r^{2}\left(d\theta^{2}+\sin^{2}\theta d\phi^{2}\right)\label{metric}
	\end{flalign}
	with
	\begin{flalign}
		f(r)=1-\dfrac{2(1+\alpha)M r^{2}}{\left(r^{2}+\alpha(1+\alpha)M^{2} \right)^{3/2}}+\dfrac{\alpha(1+\alpha)M^{2}r^{2}}{\left(r^{2}+\alpha(1+\alpha)M^{2} \right)^{2}}
	\end{flalign}
	Here $M$ is the mass parameter of the gravitating object. The associated gravi-electric field is given by
	\begin{flalign}
		E_{\rm grav}(r)=\sqrt{\alpha}Mr^{4}\left[\dfrac{r^{2}-5\alpha(1+\alpha)M^{2}}{\left\{r^{2}+\alpha(1+\alpha)M^{2}\right\}^{4}}+\dfrac{15}{2}\dfrac{(1+\alpha)M}{\left\{r^{2}+\alpha(1+\alpha)M^{2}\right\}^{7/2}} \right]
		\end{flalign}
	For a convenient way of studying the theory, we introduce the alternative parameter $\beta$ as
	\begin{flalign}
		\beta=\dfrac{\alpha}{1+\alpha}
	\end{flalign}
	The ADM mass of the gravitating object is
	\begin{flalign}
		M_{\rm ADM}=(1+\alpha)M= \dfrac{M}{1-\beta} \equiv M_{\beta}
	\end{flalign}
	We can express the metric in \ref{metric} in terms of the ADM mass with
	\begin{flalign}
		f(r)= 1-\dfrac{2M_{\beta}r^{2}}{\left( r^{2}+\beta M_{\beta}^{2}\right)^{3/2}} +\dfrac{\beta M_{\beta}^{2}r^{2}}{\left( r^{2}+\beta M_{\beta}^{2}\right)^{2}}
	\end{flalign}
	Here $M_{\beta}$ is the ADM mass of the spacetime and $\beta$ is the enhancement parameter. The gravitational source charge in terms of the ADM mass is given by
	\begin{flalign}
		Q_g=\sqrt{\beta(1-\beta)G_{N}} M_{\beta}
	\end{flalign}
	The horizon of the spacetime depends on the zeros of the function $f(r)$ and that can be used to determine the critical value of the parameter $\beta$. Let us assume $\dfrac{r^{2}}{M_{\beta}^{2}}+\beta=x^{2}$, then zeros of $f(r)$ can be determined by the equation
	
	\begin{flalign}
		a x^{4}+ b x^{3}+c x^{2}+ d x +e=0
	\end{flalign}
	where, $a=1, b=-2, c= \beta, d=2\beta, e=-\beta^{2}$. The discriminant of the quartic equation is
	\begin{flalign}
		\Delta=& 256 a^{3} e^{3}-192 a^{2} b d e^{2}-128 a^{2}c^{2} e^{2}+144 a^{2}c d^{2}e-27 a^{2} d^{4}\nonumber\\
		& +144ab^{2}c e^{2}-6 a b^{2} d^{2} e-80 abc^{2} d e+18a b c d^{3}+16a c^{4}e\nonumber\\
		& -4a c^{3} d^{2} - 27 b^{4} e^{2}+18 b^{3}c d e -4 b^{3} d^{3}-4 b^{2}c^{3}e +b^{2}c^{2} d^{3}\nonumber\\
		=&-4\beta^{2}\left( 27+4\beta\left[-16+\beta\left\{20+\beta(-28+25\beta)\right\}\right]\right)
	\end{flalign}
	As the discriminant satisfies $\Delta \leq 0$, there will be two distinct real roots of the quartic equation. The critical value at which there will be only one horizon is given by the solution of the equation
	\begin{flalign}
		10800 \beta ^3-12096 \beta ^2+20304 \beta -6912=0
	\end{flalign}
	The solution of this equation is $\beta=0.402186=\beta_{\rm crit}$. For $\beta<\beta_{\rm crit}$ there will be a black hole with two horizons\cite{Moffat:2014aja} and there will be no horizon for $\beta>\beta_{\rm crit}$. This result is displayed in \ref{Zero_func}.
	
	\subsection{Motion of photons in MOG spherically symmetric spacetime}
	
\begin{figure}[h!]
		\centering
		\subfloat[Minima of the potential have been shown here. This corresponds to stable circular orbits. However, the relevance of the stable circular orbit is valid only for $\beta_{\rm crit}<\beta\lesssim 0.5$ ]
		{{\includegraphics[width=8cm]{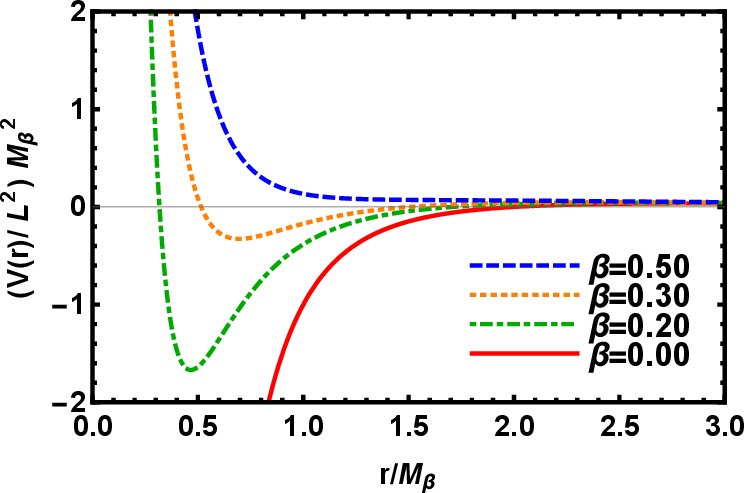} } \label{}}
		\qquad
		\subfloat[Maxima of the potential have been shown here.  This corresponds to the existence of the unstable circular orbits and is valid for the range $0<\beta\lesssim 0.5$ ]
		{{\includegraphics[width=8.2cm]{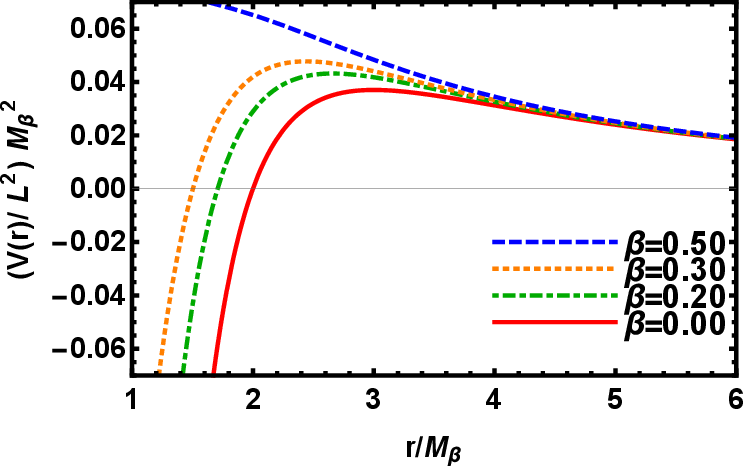} } \label{}}
		\caption{The variation of the potential for the photon particle has been shown in these plots (a) minima of the potential has been shown and (b) maxima of the potential has been shown . For $\beta<\beta_{\rm crit}$, there exists a stable circular orbit along with an unstable circular orbit. However, for the MOG static spherically symmetric solution only unstable circular orbits exist for $\beta<\beta_{\rm crit}$. The existence of both stable and unstable circular orbits is possible for $\beta_{\rm crit}<\beta \lesssim 0.5$.}
		\label{Pot}
	\end{figure}

\begin{figure}[h!]
		\centering
		\subfloat[Variation of various radii with variation of the parameter $\beta$. ]
		{{\includegraphics[width=8cm]{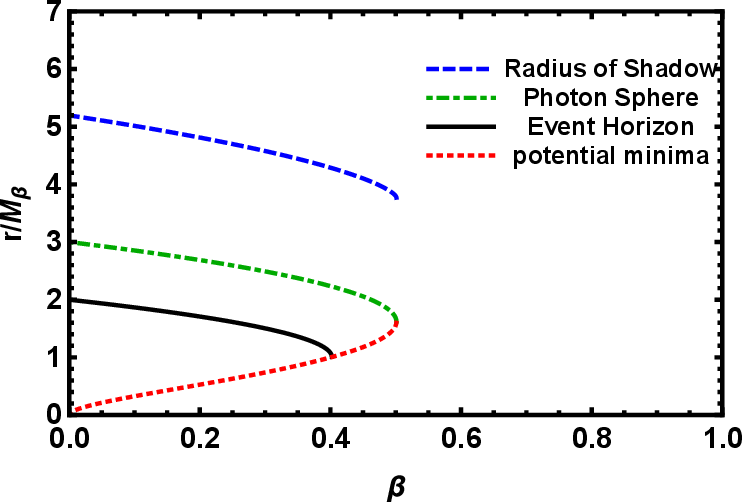} } \label{F1a}}
		\qquad
		\subfloat[Variation of the shadow radius with variation of parameter $\beta$.]
		{{\includegraphics[scale=0.6]{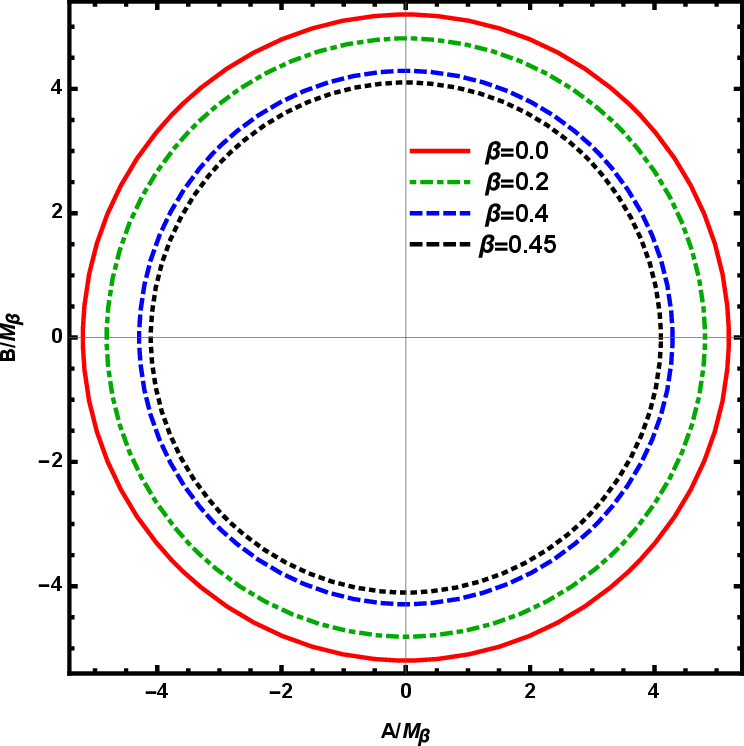} } \label{F1b}}
		
		\caption{(a)The variation of the horizon radius, photon sphere and the radius of the shadow are depicted as a function of the parameter $\beta$. It is interesting to note that in the range $0.4\lesssim \beta \lesssim 0.5$ there is no event horizon. However, this does not hinder us in defining the photon sphere and shadow for the compact object. Also, for the range of parameter space, we have both stable and unstable circular orbits. (b) The circular shadow structures have been depicted.}
		\label{}
	\end{figure}

	The Lagrangian for the photon motion is given by
	\begin{flalign}
		\mathcal{L}=\dfrac{1}{2}\left[-f(r) \dot{t}^{2}+\dfrac{1}{f(r)}\dot{r}^{2}+r^{2}\dot{\theta}^{2}+r^{2}\sin^{2}\theta \dot{\phi}^{2}\right]
	\end{flalign}
	For a spherically symmetric spacetime, we can always choose without loss of generality $\theta=\pi/2$ and $\dot{\theta}=0$. The equations for $\dot{t}$ and $\dot{\phi}$ can be deduced using the symmetries of the MOG regular, static spherically symmetric spacetime. The associated equations are
	\begin{flalign}
		f(r)\,\dot{t}&=E\\
		r^{2}\dot{\phi}&=L
	\end{flalign}
	where $E$ and $L$ are, respectively, the energy and angular momentum of the photon. The radial equation can be written as
	\begin{flalign}
		\dot{r}^{2}+V(r)=E^{2}
	\end{flalign}
	where, $V(r)=L^{2}\dfrac{f(r)}{r^{2}}$. The structure of the potential helps to determine the presence of stable or/and unstable circular orbits. From \ref{Pot}, we conclude that for the whole parameter space, there exists a stable circular orbit and an unstable circular orbit. This special situation arises for $\beta_{\rm crit}< \beta  \lesssim 0.5$. However, with close inspection and from \ref{F1a} for $\beta<\beta_{\rm crit}$, we only have unstable circular orbits.
	
	As assumed earlier, $r^{2}/M_{\beta}^{2}+\beta=x^{2}$ can be used to find the position of the photon sphere. The position of the photon sphere can be found by the real greatest solution of the following equation
	\begin{flalign}
		x^{4}-3x^{3}+2\beta x^{2}+5 \beta x -3 \beta^{2}=0
	\end{flalign}
	For  $\beta<\beta_{\rm crit}$, there will be a photon sphere and a gradual increase in the enhancement parameter $\beta$ causes the decrease of both the horizon and shadow radius. For the dark compact object with $\beta_{\rm crit}<\beta \lesssim 0.5$, although there is no horizon, we still have a photon sphere.  In spherically symmetric spacetimes, the shadow of the black hole is circular in structure. For the regular MOG black holes, the shadow has been shown with the variation in the parameter $\beta$ in \ref{F1b}. In the figures, $A$ and $B$ are the celestial coordinates.
	
	\subsection{Parameter estimation using M87* and Sgr A* data}
	
	Although astrophysical black holes are rotating in nature, for a first-hand estimation of black hole parameters, one can use the shadow of the spherically symmetric black holes. As the shadow for spherically symmetric black holes does not depend on the inclination angle to obtain the initial estimation of the parameter $\beta$, we can work with the shadow of the regular MOG black hole solution. Apart from this, the observed shadows for M87* and Sgr A* are more or less circular in nature. This motivates us to find the observational signatures of the regular MOG black hole or compact objects in M87* and Sgr A* using EHT data.
	
	We have calculated how the radius of the photon sphere and the shadow affected the parameter $\beta$ in the previous section. We can determine the values of $\beta$ based on the size of the  angular diameter,{ which is defined as 
	\begin{flalign}
	\tan \alpha \approx \alpha =\dfrac{r_{sh}}{D}
	\end{flalign} 
Where	$r_{sh}$ is the radius of the black hole shadow, $D$ is the distance of the centre of the black hole from the observer, $2\alpha$ is the angular diameter. As the distance between the black hole and the observer is much greater than the radius of the black hole shadow, the small angle approximation is justified. }

\begin{figure}[h!]
		\centering
		
		\subfloat[ Angular diameter versus $\beta$ with observed values $(42\pm 3) \mu as$ marked in grey]
		{{\includegraphics[width=9cm]{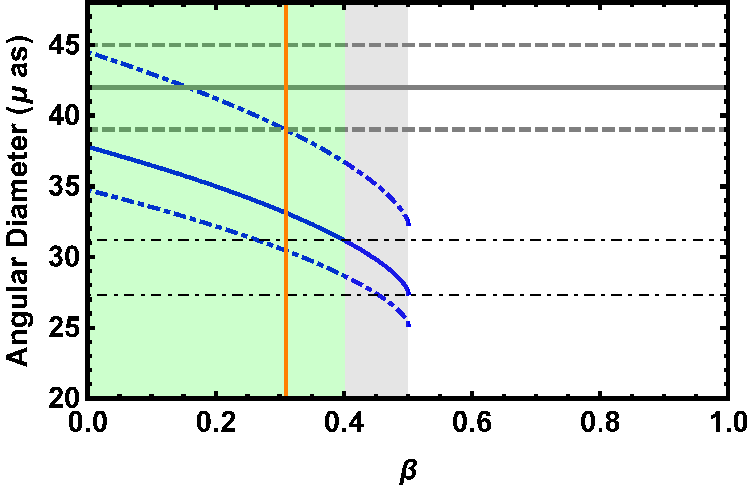} } \label{F2a}}
		\qquad
		\subfloat[ Angular diameter versus $\beta$ with observed values $(37.8\pm 2.7) \mu as$ marked in grey]
		{{\includegraphics[width=9cm]{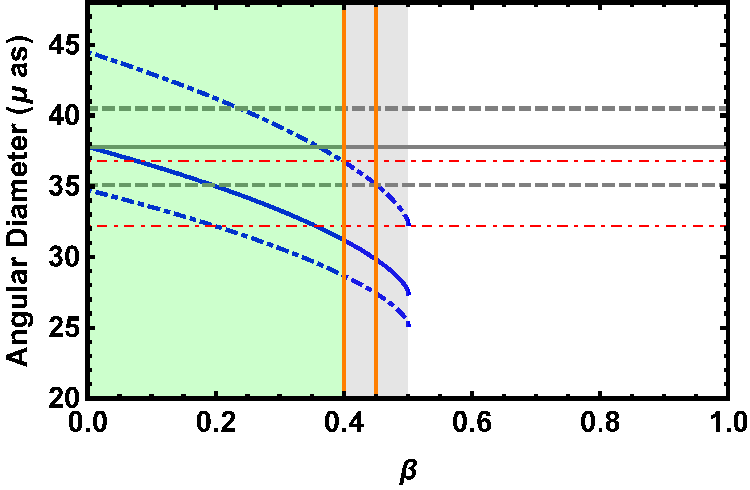} } \label{F2b}}

		\caption{(a)The variation of the ring diameter has been shown with the error bar. (b) The variation of the shadow diameter has been shown.}
		\label{Fig2}
	\end{figure}

	The mass and distance of M87* needs to be independently measured. The mass of M87* has been reported to be $M=3.5_{-0.3}^{+0.9}\times 10^{9} M_{\odot}$  from model gas dynamics mass measurements\cite{2013W}. However, based on model stellar dynamics mass measurements, the mass is reported to be $M=6.2_{-0.5}^{+1.1}\times 10^{9} M_{\odot}$\cite{Gebhardt_2009,McConnell_2011}. The distance of the gravitating source is reported to be $D=(16.8\pm 0.8) \rm Mpc$. {Having the information of mass and distance of the black hole one can define the angular gravitational radius $\theta={GM}/{c^{2}D}$. The angular gravitational radius $\theta_{dyn}$ as measured by stellar-dynamics process and the angular gravitational radius $\theta_{g}$ as reported by EHT are more or less consitent\cite{EventHorizonTelescope:2021dqv}.}
{Theoretical bounds on the shadow diameter has been discussed by Kocherlakota et. al. \cite{PhysRevD.103.104047} Based on M87* shadow size they have implied restrictions on the physical charges of several different spinning or non-rotating black holes. We use the stellar dynamics mass measurement to theoretically deduce the shadow radius of the black hole. The supermassive black hole M87* in the core of the galaxy M87 has an angular diameter of $(42\pm3)\mu as$, according to the Event Horizon Telescope (EHT) collaboration\cite{EventHorizonTelescope:2019dse}.}	In the plots shown in \ref{Fig2}, the central value $42~\mu as$ second has been shown with a grey line and the error bar has been shown with the dashed grey line. There is an error in the mass estimation of M87* around the central value. The variation of angular diameter, taking the central value of mass, has been shown with a blue line. Taking the errors, we can also plot the angular diameter. This has been shown with dot-dashed blue lines. The central value of mass of M87* is $6.2 \times 10^{9}M_{\odot}$.
	
	Considering the error bars both for angular diameter measurement and mass measurement, there is a possibility that M87* could be a regular MOG black hole. For the angular diameter $(42 \pm 3 )\mu as$ the value of the parameter $\beta$ can be as high as approximately $\beta=0.3$. This has been shown with a vertical orange line in \ref{F2a}. So, in this case we can say that the M87* is a regular MOG black hole. With the angular diameter $(42 \pm 3 )\mu as$, the possibility that M87* is a horizonless compact object can be rejected. However, if one considers a $10\%$ offset value of the angular diameter, the parameter $\beta$ can be as high as approximately $\beta=0.45$, and in this case M87* can be a horizonless compact object. In \ref{F2b}, the theoretical range of $\beta$ has been shown by a grey shaded region and the region enclosed by the two orange lines in grey shaded represents the observationally allowed range of the parameter $\beta$ for M87*.

\begin{figure}[H]
		\centering
		\includegraphics[width=9cm]{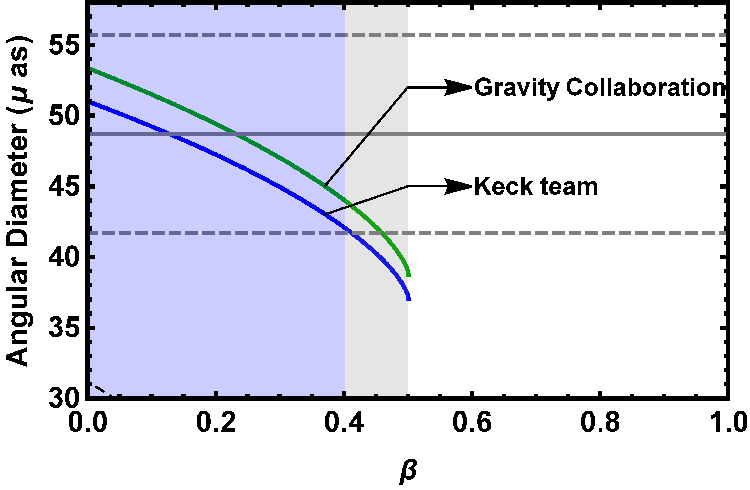}
		\caption{Theoretical angular diameter for Sgr  A* has been shown.}
		\label{AD_SGR}
	\end{figure}

	According to the EHT collaboration, the angular diameter of the Sgr A* shadow is $(48.7\pm 7)\mu as$\cite{EventHorizonTelescope:2022wkp,EventHorizonTelescope:2022vjs,EventHorizonTelescope:2022wok,EventHorizonTelescope:2022exc,EventHorizonTelescope:2022urf,EventHorizonTelescope:2022xqj}. The angular diameter of the Sgr A* shadow depends on the determined mass and distance of Sgr A*. Several groups have reported the mass and distance of Sgr A*. From the Keck team, keeping the redshift parameter free the mass and distance of Sgr A* have been reported to $(3.975\pm0.058\pm 0.026)\times 10^{6} M_{\odot}$ and $(7959\pm 59\pm 32)\rm pc$, respectively \cite{Do:2019txf}. The same group has also reported the mass and distance assuming the redshift parameter to be unity and these are $(3.951\pm 0.047)\times 10^{6} M_{\odot}$ and $(7935\pm 50)\rm pc$, respectively\cite{Do:2019txf}.  The mass and distance, according to the Gravity collaboration are, respectively, $(4.261\pm 0.012)\times 10^{6} M_{\odot}$ and $(8246.7\pm 9.3)\rm pc$\cite{GRAVITY:2021xju,GRAVITY:2020gka}. The Gravity Collaboration further limited the BH mass $(4.297\pm 0.012\pm 0.040)\times 10^{6} M_{\odot}$ and the distance $(8277\pm 9 \pm 33)\rm pc$ by accounting for optical aberrations. In \ref{AD_SGR}, we have plotted the angular diameter as a function of the parameter $\beta$ with mass and distance as given by the above teams. From the plot, using the Keck team data, one can constrain the parameter to be $0<\beta\lesssim 0.4$. With the Keck team data, it is almost impossible to say that Sgr A* is a horizonless compact object. However, using the Gravity collaboration data, we can say that there is a possibility that Sgr A* is a horizonless compact object, because with the Gravity collaboration data the parameter range is $0<\beta \lesssim 0.46$.
	
	\section{Regular MOG rotating compact object}\label{Sec-RS}
	
\begin{figure}[h!]
		\centering
		\includegraphics[width=9cm]{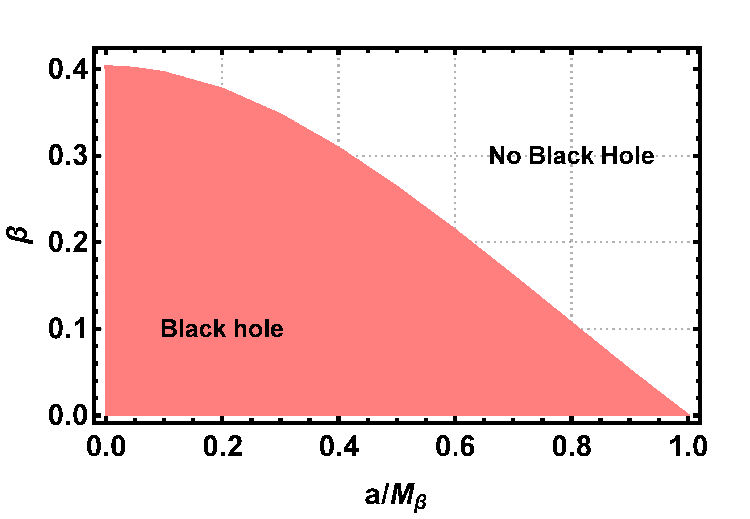}
		\caption{Parameter space of $(\beta-a)$ plane for the regular rotating MOG solution is displayed here. The reddish region represents the black hole solution and the boundary denotes the occurrence of an extremal black hole.  }\label{pspace}
	\end{figure}

	The regular rotating MOG solution can be obtained with the help of the modified Newman-Janis algorithm. The associated line element of the spacetime in Boyer-Lindquist coordinates is given by\cite{Moffat:2018jmi}
	\begin{flalign}
		ds^{2}=& - f(r,\theta) dt^{2}-2a\sin^{2}\theta\left\{ 1-f(r,\theta)\right\}d\phi dt  \nonumber\\
		& +\dfrac{\Sigma}{\Delta}dr^{2}+ \Sigma d\theta^{2} + \sin^{2}\theta \left[\Sigma -a^{2}\left\{f(r,\theta)-2 \right\}\sin^{2}\theta \right] d\phi^{2}
		\label{M_2}
	\end{flalign}
	where
	\begin{subequations}
		\begin{flalign}
			f(r,\theta)
			&=1-\dfrac{2M_{\beta}r\sqrt{\Sigma}}{\left[\Sigma+\beta M_{\beta}^{2} \right]^{3/2}} + \dfrac{\beta M_{\beta}^{2}\Sigma}{\left[\Sigma+\beta M_{\beta}^{2} \right]^{2}}\\
			\Delta & =\Sigma f(r,\theta)+a^{2}\sin^{2}\theta\\
			\Sigma & = r^{2}+a^{2}\cos^{2}\theta
		\end{flalign}
	\end{subequations}
	Here, $M_{\beta}$ is the ADM mass of the spacetime, $\beta$ is the enhancement parameter and $a$ is the spin parameter.
	A certain portion of the full parameter space of $\beta-a$ is available for the existence of the regular rotating MOG black hole. The parameter space has been shown in \ref{pspace}. From the figure, it is noticeable that the highly spinning regular MOG black hole has a relatively low value of the parameter $\beta$.

\begin{figure}[h!]
		\centering
		\subfloat[ Variation of $g_{tt}$ with respect to $r$, when $\theta=0$ and $\beta=0.1$]
		{{\includegraphics[width=7 cm]{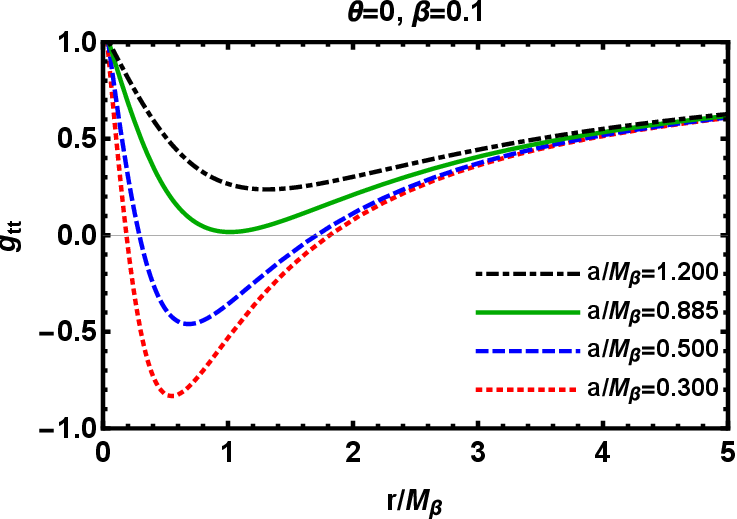} } }
		\qquad
		\subfloat[Variation of $g_{tt}$ with respect to $r$, when $\theta=0$ and $\beta=0.3$ ]
		{{\includegraphics[width=7cm]{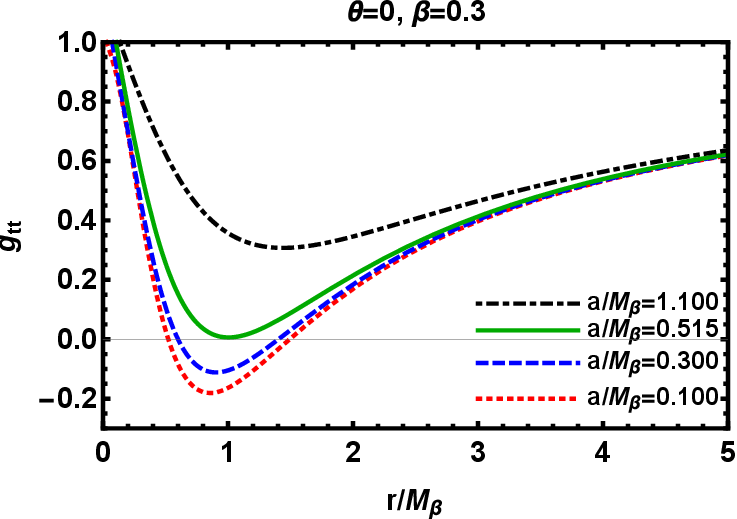} } }
		\qquad
		\subfloat[Variation of $g_{tt}$ with respect to $r$, when $\theta=\pi/4$ and $\beta=0.1$ ]
		{{\includegraphics[width=7cm]{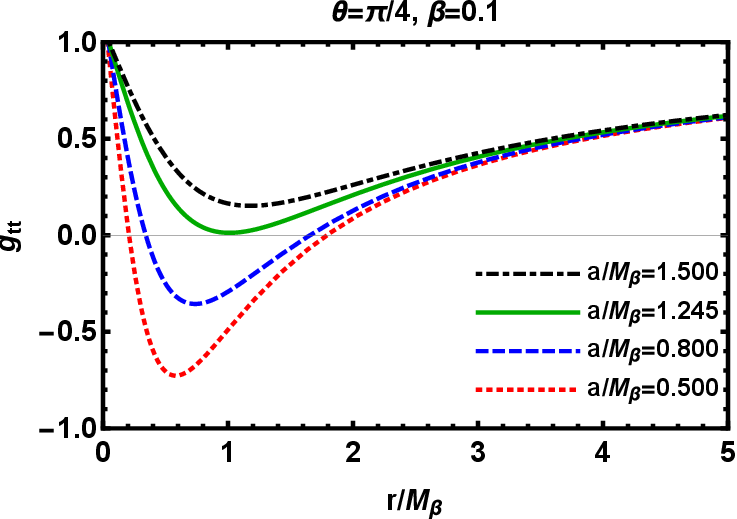} } }
		\qquad
		\subfloat[Variation of $g_{tt}$ with respect to $r$, when $\theta=\pi/4$ and $\beta=0.3$ ]
		{{\includegraphics[width=7cm]{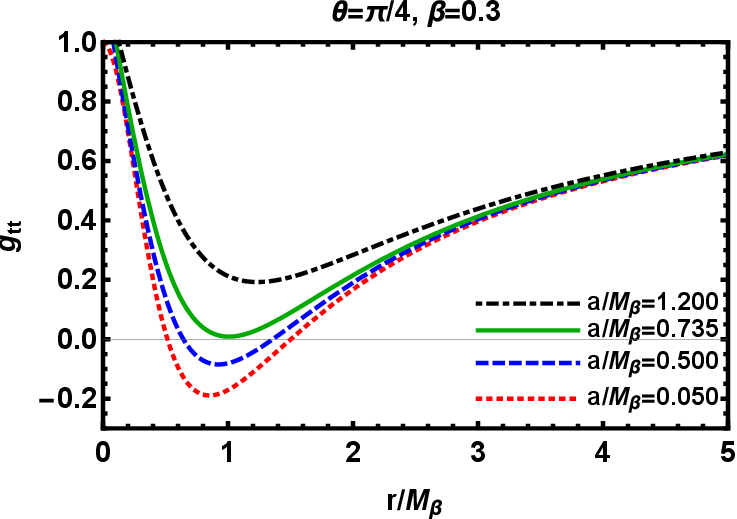} } }

		\caption{The nature of the static limit surface (SLS) has been depicted as a function of coordinate $r$ for various values of spin parameter $a$. The variation of the SLS with respect to the enhancement parameter $\beta$ can be seen along the row of the figure matrix [i.e (a)-(b) and (c)-(d)]}
		\label{SLS}
	\end{figure}

	The location and the structure of the static limit surface are obtained by setting the prefactor of $dt^{2}$ to zero. The SLS can be determined by solving the following equation
	\begin{flalign}
		\left(\Sigma + \beta M_{\beta}^{2} \right)^{2} -2 M_{\beta} r \sqrt{\Sigma (\Sigma +\beta M_{\beta}^{2})} + \beta M_{\beta}^{2}\Sigma =0
	\end{flalign}
	For $\beta=0$, we have the usual Kerr scenario. The variation and existence of the static limit surface for rotating regular MOG black holes has been displayed in \ref{SLS}.
	The location of the horizon is determined by setting the $g^{rr}$ to be zero, which in turn gives
	\begin{flalign}
		\Delta=\Sigma f(r,\theta)+a^{2}\sin^{2}\theta=0
	\end{flalign}
	\begin{figure}
		\begin{center}
		\end{center}
	\end{figure}
	The existence of the horizon in the regular rotating MOG solution has been depicted in \ref{grr}.
	
	\begin{figure}[h!]
		\centering
		\subfloat[Variation of $g^{rr}$ with respect to $r$ when $\beta=0.1$]
		{{\includegraphics[width=7cm]{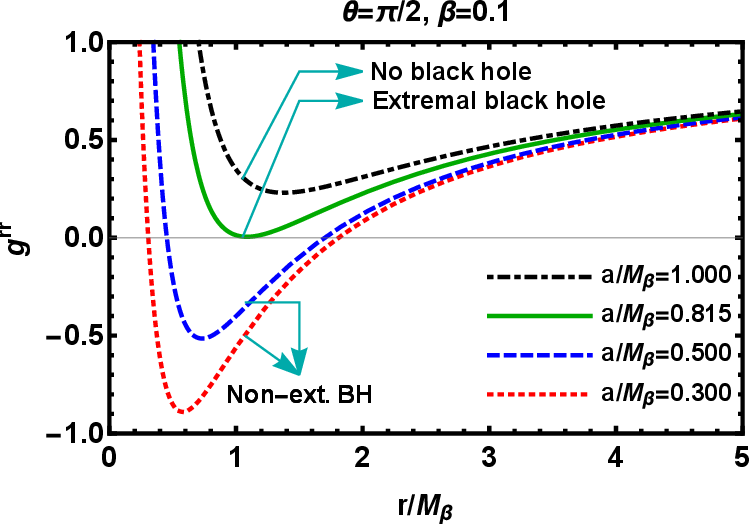} }}
		\qquad
		\subfloat[Variation of $g^{rr}$ with respect to $r$, when  $\beta=0.3$ ]
		{{\includegraphics[width=7cm]{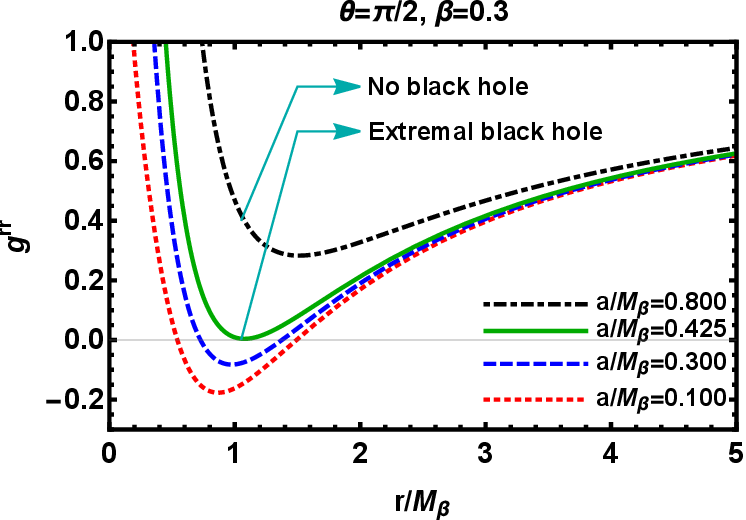} }}
		\caption{A set of parameter values are allowed for black hole solutions. In these plots, the set of parameters has been shown for the extremal black hole.  The   horizonless compact object can result for an increase in spin keeping the enhancement parameter $\beta$ constant. (a) The enhancement parameter $\beta$ is $0.1$. Whereas in (b) the enhancement parameter is $0.3$.}\label{grr}
	\end{figure}.

	\section{Analysis of the black hole shadow}\label{Sec-NG}
	
	To investigate the black hole shadow, we must determine the photon geodesic equations for the metric given in \ref{M_2}. When using the Hamilton-Jacobi formulation for the rotating MOG regular solution, it is particularly challenging to separate the equations, since the function $f(r,\theta)$ has a highly complex structure. Consequently, to overcome this issue, we consider an approximation for $\theta$, such that $\theta \approx \pi/2+\epsilon $.\cite{Abdujabbarov:2016hnw}
	Note that although we are focusing on photon orbits that are close to the equator, unstable photon circular orbits are not only limited to this region. This fact does not invalidate the calculations that follow, because the major goal of this work is to calculate the shadow of a black hole cast by an observer at infinity, which can be done using the approximations indicated above. The trigonometric functions here have the following form: $\sin\theta\approx1$ and $\cos\theta\approx-\epsilon$. With these approximations the function $f(r,\theta)$ becomes $f(r)$, which is given by
	\begin{flalign}
		f(r)= 1-\dfrac{2M_{\beta}r^{2}}{\left( r^{2}+\beta M_{\beta}^{2}\right)^{3/2}} +\dfrac{\beta M_{\beta}^{2}r^{2}}{\left( r^{2}+\beta M_{\beta}^{2}\right)^{2}}
	\end{flalign}
	\subsection{Null geodesics}
	For a general stationary, axisymmetric metric the Lagrangian $\mathcal{L}$ can be written as
	\begin{flalign}
		g_{\mu\nu}\dot{x}^{\mu}\dot{x}^{\nu}= g_{tt}\dot{t}^{2}+2g_{t\phi}\dot{t}\dot{\phi}+g_{\phi\phi}\dot{\phi}^{2}+g_{rr}\dot{r}^{2}+g_{\theta\theta}\dot{\theta}^{2}=2\mathcal{L}
		\label{gen_M}
	\end{flalign}
	For massive particles and massless particles, the Lagrangian is equal to unity and zero, respectively. The associated Hamiltonian is provided by
	\begin{flalign}
		\mathcal{H}=p_{\mu}\dot{x}^{\mu}-\mathcal{L}=\dfrac{1}{2}g^{\mu\nu}p_{\mu}p_{\nu}=\dfrac{k}{2}
		\label{H}
	\end{flalign}
	with $k$, which in this instance is zero and represents the test particle's rest mass. By utilising the Hamilton-Jacobi technique, we may connect the Hamiltonian to the action S by
	\begin{flalign}
		\mathcal{H}(x^{\mu},p^{\mu})+\dfrac{\partial S}{\partial \lambda}=0 \hspace{1cm} {\rm with} \hspace{1cm}  p_{\mu}=\dfrac{\partial S}{\partial x^{\mu}}
	\end{flalign}
	The metric described in \ref{gen_M} is independent of $t$ and $\phi$, whereby the (specific) energy E and the (specific) angular momentum L are conserved quantities. These are supplied by
	\begin{subequations}
		\begin{flalign}
			E=g_{tt}\dot{t}+g_{t\phi}\dot{\phi}\\
			L=g_{t\phi}\dot{t}+g_{\phi\phi}\dot{\phi}
		\end{flalign}
	\end{subequations}
	From the aforementioned context, the action may be expressed as
	\begin{flalign}
		S=-Et+L\phi+S(r,\epsilon)
		\label{sep}
	\end{flalign}
	Now, for the metric given in \ref{M_2}, \ref{sep} is separable such that $S(r,\theta)=S_{r}(r)+S_{\epsilon}(\epsilon)$. Substituting \ref{sep} in equation \ref{H} we get
	\begin{flalign}
		g^{rr} \left(\dfrac{\partial S^{r}}{\partial  r}\right)^{2} + g^{\theta\theta} \left(\dfrac{\partial S^{\theta}}{\partial  \theta}\right)^{2} +g^{tt}(-E)^{2}+ 2 g^{t\phi} (-E)L + g^{\phi\phi} L^{2}=0
	\end{flalign}
	The metric in \ref{M_2} causes the equation above to take the form:
	\begin{flalign}
		\Delta \left(\dfrac{dS^{r}}{dr}\right)^{2}+\left(\dfrac{dS^{\epsilon}}{d\epsilon}\right)^{2} - \left\{ \dfrac{1}{\Delta}\left[r^{2}+a^{2}\right]^{2}-a^{2}\right\} E^{2} + \dfrac{2a r^{2}}{\Delta}\left\{1-f(r)\right\} EL+ \dfrac{r^{2}f(r)}{\Delta}L^{2}=0
	\end{flalign}
	It's interesting to note that the $r$ and $\theta$ components of the preceding equation may be split up so that
	\begin{flalign}
		& \Delta \left(\dfrac{dS^{r}}{dr}\right)^{2}-\dfrac{1}{\Delta}\left[r^{2}+a^{2}\right]^{2}E^{2} + \dfrac{2a r^{2}}{\Delta}\left\{1-f(r)\right\}EL+\dfrac{r^{2}f(r)}{\Delta}L^{2}+a^{2}E^{2}   = -\left(\dfrac{dS^{\epsilon}}{d\epsilon}\right)^{2}=-C
		\label{C1}
	\end{flalign}
	The Carter constant is represented by C. The left-hand side of \ref{C1} is only a function of $r$, whereas the right-hand side is a function of $\theta$ alone. The radial component of \ref{C1} may be expressed as
	\begin{flalign}
		\left[\dfrac{dS^{r}}{dr} \right]^{2} = \dfrac{R(r)}{\Delta^{2}}
	\end{flalign}
	where
	\begin{flalign}
		R(r) = -\Delta \left[ C +(L-aE)^{2}\right]+ \left\{ \left[r^{2} +a^{2} \right]E -aL\right\}^{2}
		\label{R1}
	\end{flalign}
	The angular part can be written as
	\begin{flalign}
		\left(\dfrac{dS^{\epsilon}}{d\epsilon}\right)^{2}  = C
	\end{flalign}
	Consequently, the action adopts the form
	\begin{flalign}
		S=  -E t + L\phi + \int \dfrac{\sqrt{R(r)}}{\Delta} dr + \int \sqrt{C} d\epsilon
	\end{flalign}
	The equation of motion for $r$ and $\epsilon$ is given by
	\begin{flalign}
		\dot{r}=\pm \dfrac{\sqrt{R}}{r^{2}}\\
		\dot{\epsilon}=\pm \dfrac{\sqrt{C}}{r^{2}}
	\end{flalign}
	For determining the unstable circular orbits, one needs to introduce $\chi=\dfrac{C}{E^{2}}$ and $\eta=\dfrac{L}{E}$. The unstable circular orbit can be obtained by setting $R(r)=0=\dfrac{dR(r)}{dr}$. So, using \ref{R1} with aforementioned conditions one obtains
	\begin{flalign}
		\left[r^{2}f(r)+a^{2}\right] \left(\chi + \eta^{2}+a^{2}-2\eta a \right)=\left[r^{2}+a^{2}-a\eta\right]^{2}\\
		{\chi +\eta^{2}+a^{2}-2\eta a} = \dfrac{4}{ \left[2f(r)+rf'(r)\right]}\left[r^{2}+a^{2}-a\eta\right]
	\end{flalign}
	These two equations can be solved to get two one-parameter classes of solutions parametrized in terms of $r$, which is the radius of unstable circular orbits:
	\begin{enumerate}[(i)]
		\item
		\begin{flalign}
			\chi &= -\dfrac{r^{4}}{a^{2}}\\
			\eta &= \frac{a^2 + r^{2}}{a }
		\end{flalign}
		\item
		\begin{flalign}
			\chi&=\frac{r^3 \left[8 a^2 f'(r)-r \left\{r f'(r)-2 f(r)\right\}^2\right]}{a^2 \left\{r f'(r)+2 f(r)\right\}^2}\\
			\eta&=\dfrac{1}{a}\left[r^{2}+a^{2}-\dfrac{4(r^{2}f(r)+a^{2})}{2f(r)+rf'(r)} \right]
		\end{flalign}
	\end{enumerate}
	The solution of the first kind is not a physical solution, but the second solution helps to determine the contour of the shadow in the $(\eta,\chi)$ plane. Further, this solution satisfies the following condition for the critical curve:
	\begin{flalign}
		a^{2}-\chi -\eta^{2}=\frac{8 \left(a^2+r^2 f(r)\right)}{r f'(r)+2 f(r)}-\frac{16 \left(a^2+r^2 f(r)\right)}{\left(r f'(r)+2 f(r)\right)^2}-2 r^2
	\end{flalign}
	When we consider the non-rotating case i.e the regular MOG static spherically symmetric solution, we have
	\begin{flalign}
		\chi+\eta^{2}=\dfrac{2r_{ph}^{2}\left[4f(r_{ph})^{2}-8f(r_{ph}) -r_{ph}^{2}f'(r_{ph})^{2} \right]}{\left\{r_{ph} f'(r_{ph})+2 f(r_{ph})\right\}^2}
	\end{flalign}
	Here, $r_{ph}$ is the radius of the photon sphere. The above equation helps to find the shadow of the regular MOG static, spherically symmetric solution.

	\subsection{Celestial coordinates and shadow structure}
	
	We now want to find out how the rotating MOG regular black hole shadow appears to be shaped. For a clearer depiction, we locate the shadow using the celestial coordinates $A_{i}$ and $B_{i}$. These coordinates are introduced as
	\begin{flalign}
		A_{i}&=\displaystyle \lim_{r_{0}\to \infty} \left(-r_{0}^{2}\sin\theta_{0} \dfrac{d\phi}{dr} \right)\\
		B_{i}&=\displaystyle \lim_{r_{0}\to \infty} \left(r_{0}^{2}\dfrac{d\epsilon}{dr} \right)
	\end{flalign}
	where $r_{0}$ is the distance between the black hole and the distant observer and $\theta_{0}$ is the inclination angle i.e the angle between the line of sight and the rotation axis of the black hole. From further calculations and considering the limit, one can arrive at the following:
	\begin{flalign}
		A_{i}=-\eta\\
		B_{i}=\sqrt{\chi}
	\end{flalign}
	
	\begin{figure}[h!]
		\centering
		\includegraphics[width=12cm]{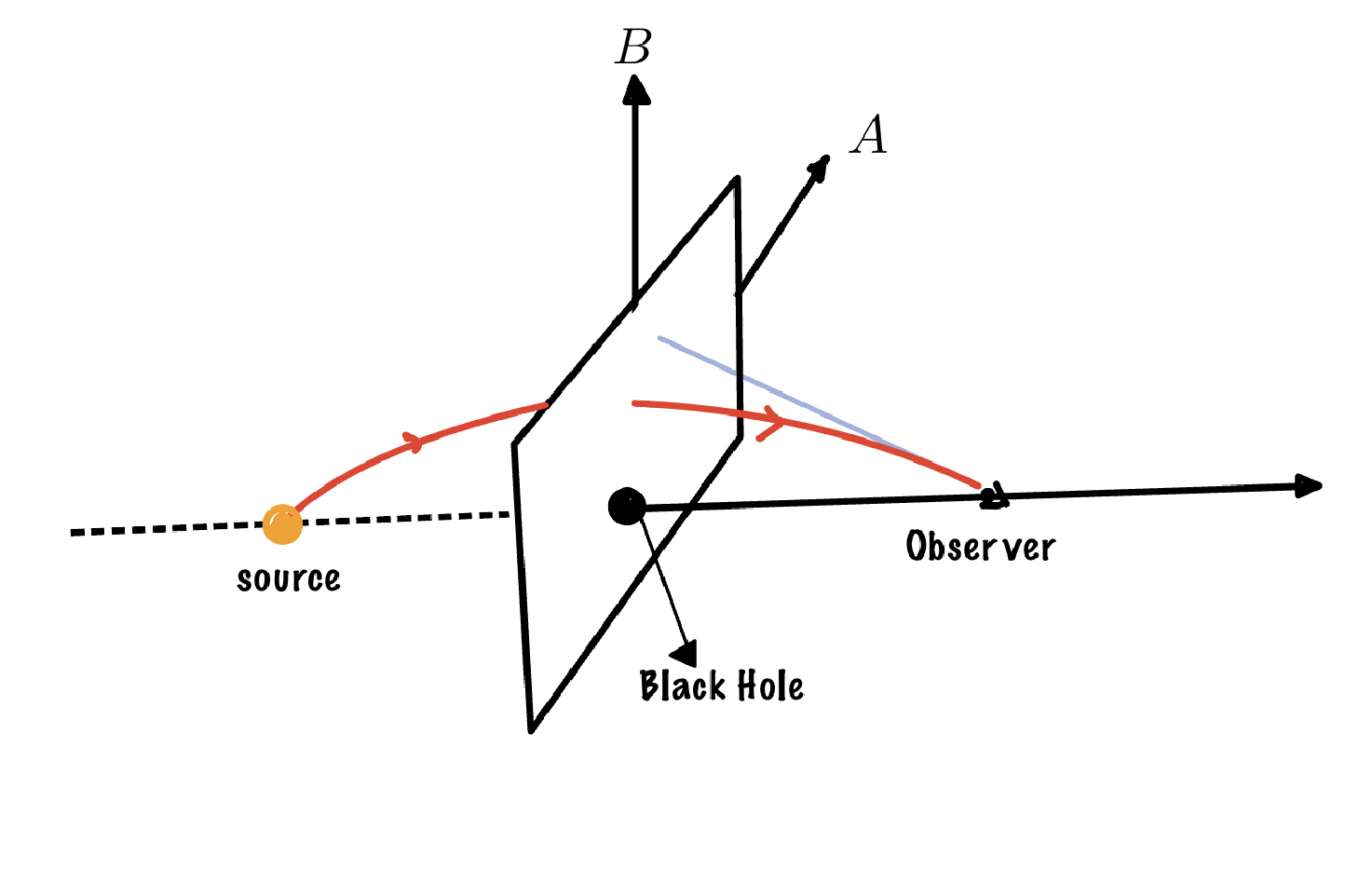}
		\caption{Schematic diagram of lensing and formation of shadow}\label{Schm}
	\end{figure}
\begin{figure}[h!]
		\centering
		\subfloat[]
		{{\includegraphics[width=6cm]{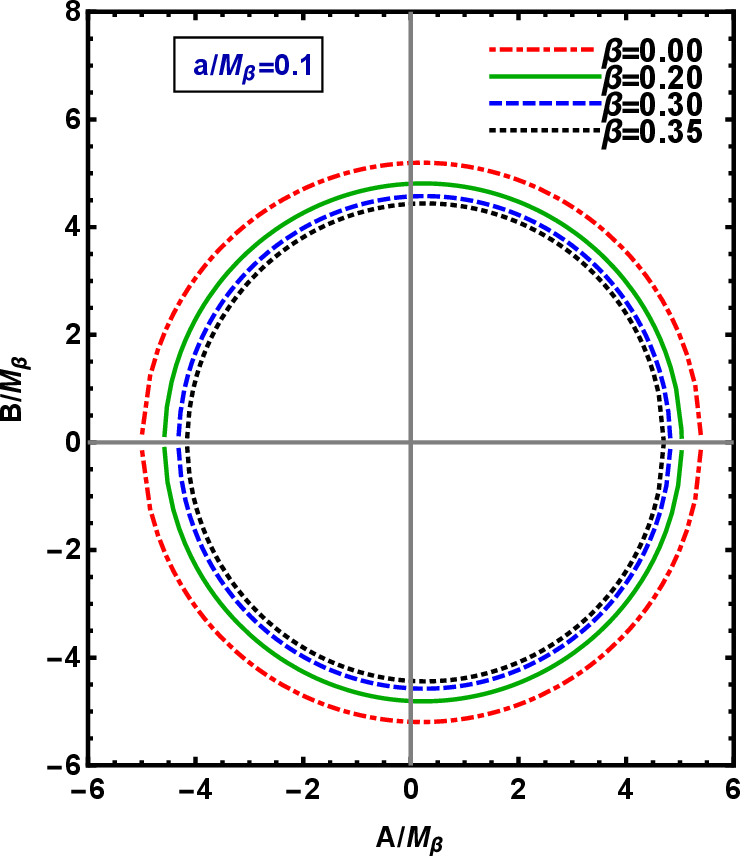} }}
		\qquad
		\subfloat[ ]
		{{\includegraphics[width=6cm]{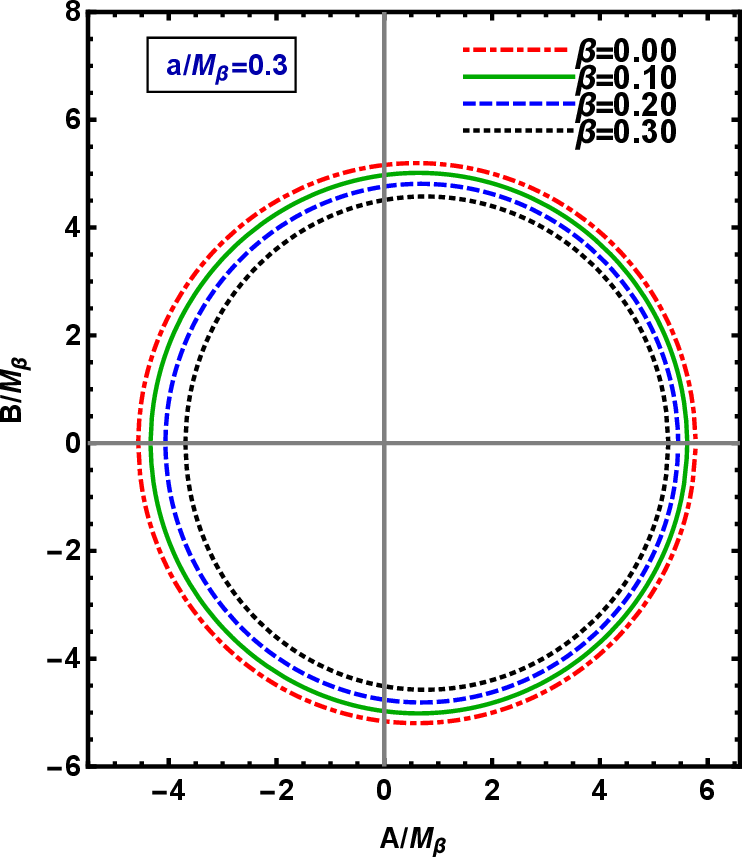} }}
		\qquad
		\subfloat[ ]
		{{\includegraphics[width=6cm]{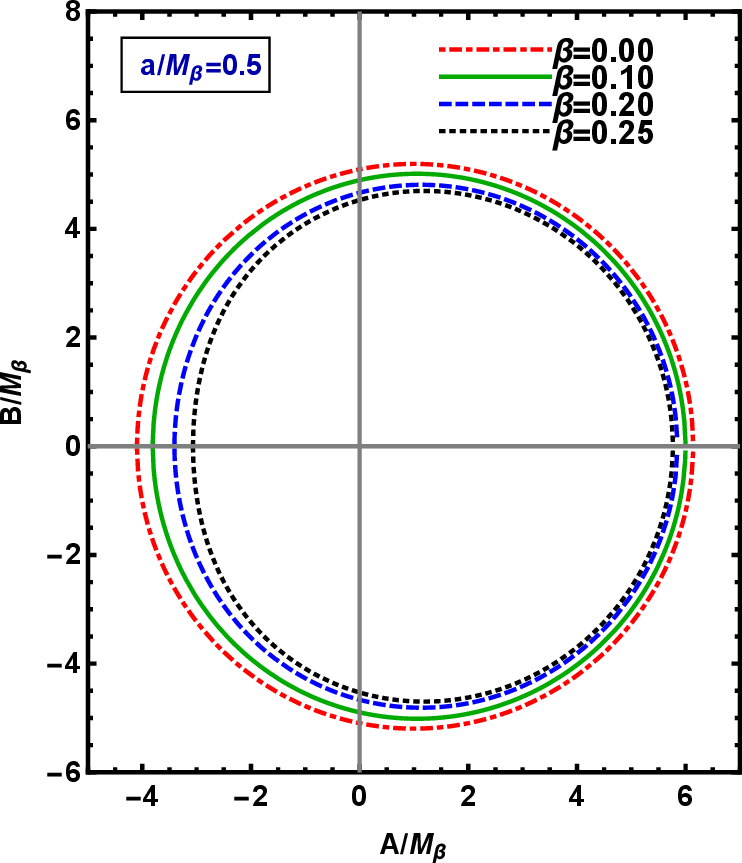} }}
		\qquad
		\subfloat[ ]
		{{\includegraphics[width=6cm]{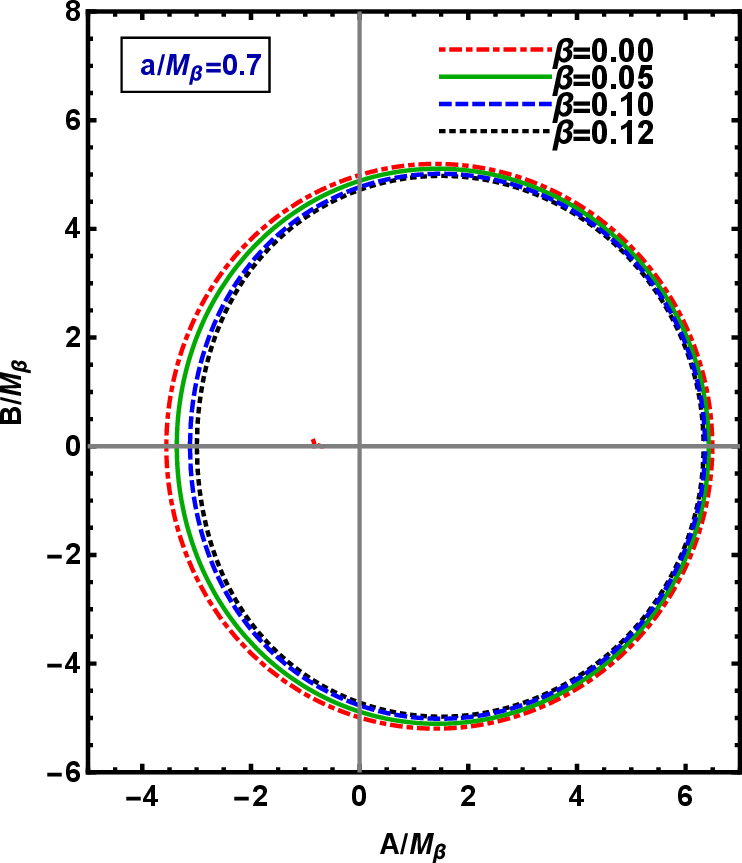} }} 	
		\caption{Shadow of the regular MOG rotating black hole is situated at the origin of the coordinate system. The inclination angle is $\theta_{0}=\pi/2$. Each image represents the shadow for a fixed value of the spin parameter a. One can note that just like in the Kerr scenario, here also the shadow gets dented with an increase in spin parameter $a$. An increase in the parameter $\beta$ causes a shrinking of the shadow for a fixed value of the spin parameter.}
		\label{shad}
	\end{figure}

	The photons are now parametrized by the conserved quantities $(\eta,\chi)$. All the light rays, coming from the source placed behind the black hole, will not be able to reach the observer as the black hole will hinder a portion of light rays due to gravitational lensing as shown in \ref{Schm}. The dark patch that appears to the observer is known as the black hole shadow and the boundary of this shadow can be determined allowing the parameters $(\eta,\chi)$ all possible values. The coordinates $A_{i}$ and $B_{i}$ are known as celestial coordinates. For a spherically symmetric scenario, these coordinates do not depend on the inclination angle and the shadow appears to be a perfect circle. However, the inclusion of the rotational effect of compact objects and the inclination angle make the shadow dented and not a perfect circle. The shape and size of the shadow can be analysed to determine the parameters including the spin parameter of the black hole. The variation of the shape and size of the black hole shadow has been depicted in \ref{shad}.

	\subsection{Observables}
\begin{figure}[h!]
		\centering
		\includegraphics[width=9cm]{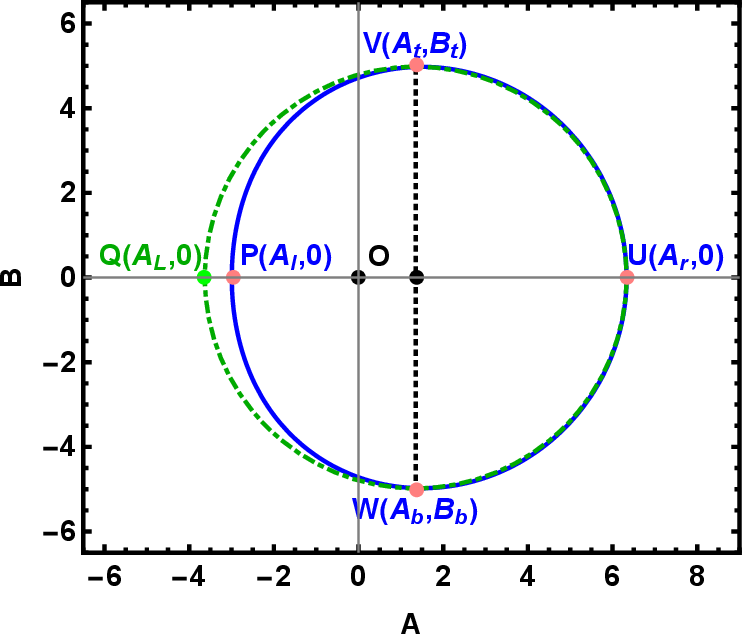}
		\caption{Characteristic points have been shown in the schematic diagram of the black hole shadow. A solid blue curve is the outline of the black hole shadow. The dot-dashed green curve is associated with the fitting circle. The associated circle passes through the three points of the shadow outline. The top-most and bottom-most points of the shadow are, respectively, $V$ and $W$. The left-most and right-most points of the shadow outline are $P$ and $U$. A separation distance of points $P$ and $Q$ measures the distortion parameter $\delta_{s}$.}
		\label{Outline}
	\end{figure}

\begin{figure}[h!]
		\centering
		\subfloat[Variation of $R_{s}$ with $\beta$]
		{{\includegraphics[width=10cm]{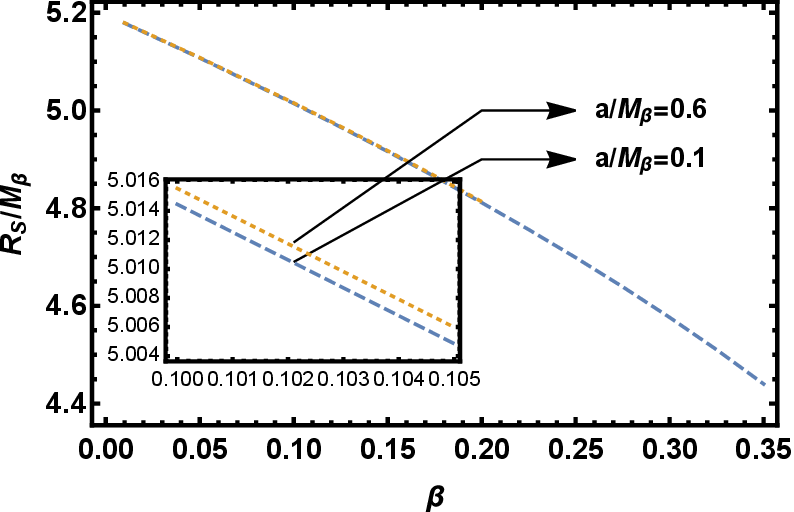} }}
		\qquad
		\subfloat[Variation of $\delta_{s}$ with $\beta$ ]
		{{\includegraphics[width=10cm]{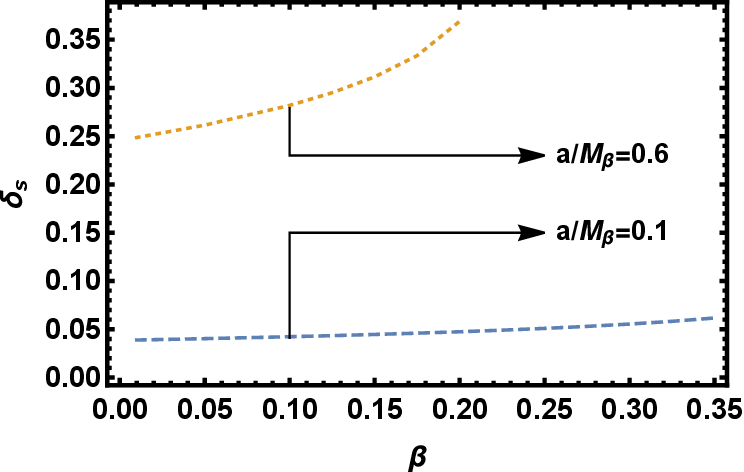} }}
		\caption{The variation of characteristic radius $R_{s}$ and distortion parameter $\delta_{s}$ of the regular rotating MOG black hole as a function of parameter $\beta$ has been shown.  The variation has been shown for a fixed value of spin parameter $a$. From the plots, one notices how the observables get changed with the spin parameter.}
		\label{RSDS}
	\end{figure}

	For further analysis of the shape of the critical curve, we are going to define two new observables as prescribed by Hioki and Maeda \cite{Hioki:2009na}. To define these observables, we need to characterize a few points of the critical curve while fitting it with a circular outline. Consider a circle in \ref{Outline} that passes through the three extreme points of the shadow curve. The points are:
	\begin{itemize}
		\item extreme right of the shadow i.e $U(A_{r},0)$, at which the shadow intersects the $A-\rm axis$.
		\item top-most point of the shadow i.e $V(A_{t},B_{t})$
		\item bottom-most point of the shadow i.e  $W(A_{b},B_{b})$
	\end{itemize}
	As the shape of the shadow is not circular, the extreme left point of the shadow does not coincide with the extreme left point of the associated circle. This characterizes the distortion of the shape of the shadow from a circular shape. The extreme left point of the shadow is $P(A_{l},0)$ and the extreme point of the associated circle is $Q(A_{L},0)$. Now we define the two observables associated with the shadow curve, which are
	\begin{enumerate}[(i)]
		\item The characteristic radius $R_{s}$, which can be defined as
		\begin{flalign}
			R_{s}=\dfrac{(A_{t}-A_{r})^{2}+B_{t}^{2}}{2 |A_{t}-A_{r}|}
		\end{flalign}
		\item Distortion parameter $\delta_{s}$, which is defined as
		\begin{flalign}
			\delta_{s}=\dfrac{D_{s}}{R_{s}}=\dfrac{|A_{L}-A_{l}|}{R_{S}}
		\end{flalign}
	\end{enumerate}
	For a non-rotating scenario, the distortion parameter becomes zero as the shape of the shadow for such a case is always zero. Similarly, the characteristic radius reduces to the radius of the circle for the non-rotating scenario. So these two observables measure the deviation from the circularity of the shape of the shadow.
	It has been illustrated in \ref{RSDS} how the parameter $\beta$ affects the characteristic radius $R_{s}$ and distortion parameter $\delta_{s}$ of the spinning regular MOG black hole.
	The change in the observables has been shown for two fixed values of the spin parameters $a$.
	
	\section{Energy emission rate for the rotating regular MOG black hole}\label{Sec-EM}
	
\begin{figure}[h!]
		\centering
		\subfloat[Energy emission rate for the value of spin parameter $a/M_{\beta}=0.20$]
		{{\includegraphics[width=10cm]{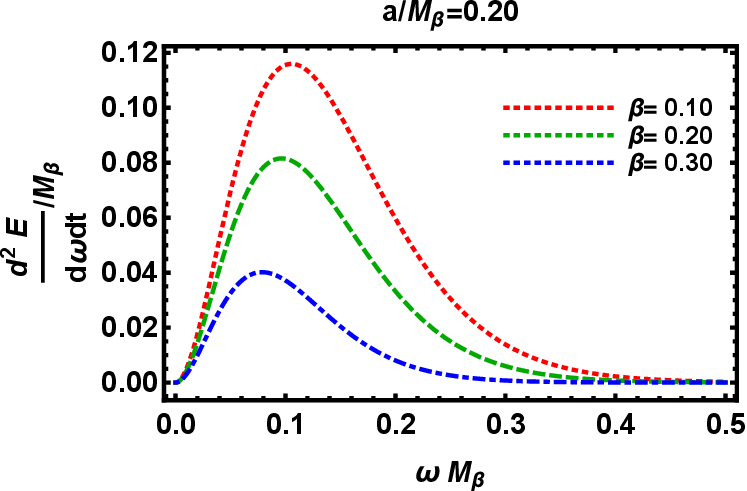} }}
		\qquad
		\subfloat[Energy emission rate for the value of the spin parameter $a/M_{\beta}=0.80$]
		{{\includegraphics[width=10cm]{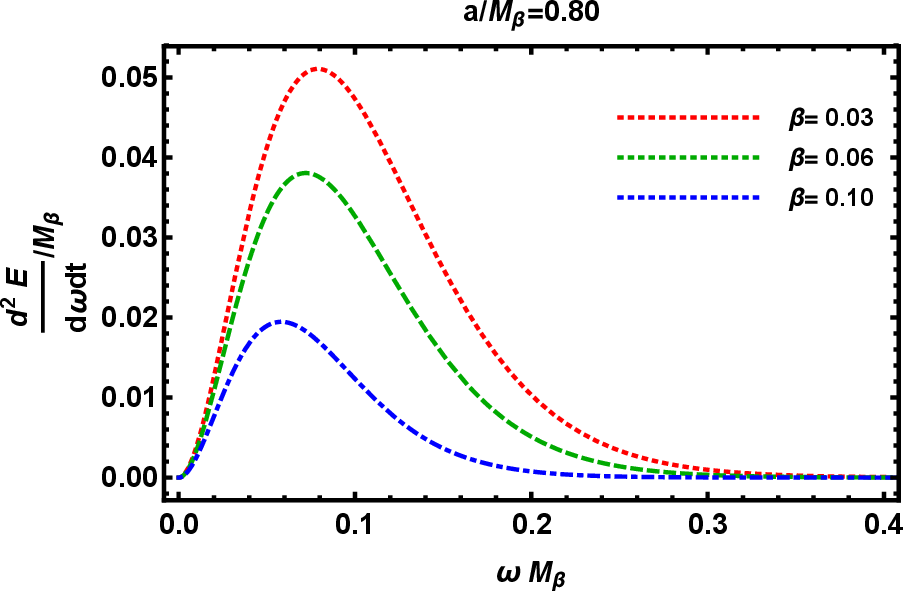} }}
		\caption{Energy emission rate as a function of frequency has been shown. For a fixed value of the spin parameter, the increase in the parameter $\beta$ causes a shift of the peak of the spectrum to a lower frequency.}
		\label{Spectrum}
	\end{figure}

	For the regular rotating MOG solution, the observers see the large energy absorption cross-section is caused by the shadows of black holes. At high energies, the black hole absorption cross sections exhibit a little modulation close to a limiting constant value. We may use the absorption cross-section limiting constant value for a nearly spherically symmetric black hole as a decent approximation, which is given by
	\begin{flalign}
		\sigma_{lim}\approx \pi R_{s}^{2}
	\end{flalign}
	The energy emission rate of the concerned black hole is given as \cite{Ma:2020dhv}:
	\begin{flalign}
		\dfrac{d^{2}E(\omega)}{d\omega dt}=\dfrac{2\pi^{3}R_{s}^{2}\omega^{3}}{e^{\omega/T}-1}
	\end{flalign}
	where $\omega$ is the frequency of the photon and $T$ is the Hawking temperature, which can be defined as \cite{Ma:2020dhv}:
	\begin{flalign}
		T=\displaystyle \lim_{\theta=0,r\to r_{+}}\dfrac{\partial_{r}\sqrt{-g_{tt}}}{2\pi\sqrt{g_{rr}}}
	\end{flalign}
	Here, $r_{+}$ is the outer event horizon of the regular rotating MOG black hole. In \ref{Spectrum}, we have plotted the energy emission rate of the black hole with the frequency of the photon $\omega$ for different values of the parameter $\beta$. One notices from the figure that the peak of the energy emission rate shifts towards a lower frequency as the parameter $\beta$ increases.
	
	\section{Conclusions}
	
	In this paper, we have explored the regular black hole solution in STVG/MOG theory. In early papers on this theory, the parameter space is from zero to infinity. We have compactified the range of the parameter space with a modification of the form of the parameter $\beta$.  At first, we have focused on the regular solution of STVG/ MOG theory in the static, spherically symmetric scenario and also determined analytically the critical value of the parameter $\beta$. For the dimensionless parameter $\beta\lesssim 0.4$, we have a black hole solution with two horizons in the spherically symmetric case. For $0.4\lesssim \beta \lesssim 0.5$, there is no black hole solution as there exists no horizon. For the critical value of the parameter $\beta\cong 0.40263$, a single horizon black hole solution can be obtained. We have also studied the null geodesics in this spacetime as it is a prerequisite to analyse the shadow of the black hole. For $\beta< \beta_{\rm crit}$, only unstable circular orbits exist. However, for $\beta_{\rm crit}< \beta \lesssim 0.5$, there exists a stable circular orbit. It is also noticeable that the radius of the photon sphere, the radii of the shadow and the event horizon decrease as the parameter $\beta$ increases. Thus, the circular shadow shrinks as the parameter $\beta$ is increased. Furthermore, as the shadows of M87* and Sgr A* are more or less circular in shape, we have tried to compare the theoretical outcomes with the observational data. For this purpose, we have used independent mass measurements to calculate the theoretical angular diameter. The regular MOG black holes and the possibility of horizonless compact objects are compatible with the EHT data and mass measurements.
	
	We have considered the regular rotating MOG black hole by studying the behaviours of the horizon and static limit surface for a change in the
	parameter $\beta$. Just like the spherically symmetric case, here also a critical value of parameter $\beta$ exists for a fixed value of the spin parameter $a$. We have determined the parameter space for which a rotating regular black hole exists. As a chief goal of this paper, a special emphasis has been placed on the black hole shadow. However, we have considered only equatorial approximations. How the shape and size of the associated shadow transforms have been determined. As the spin parameter, $a$, increases the shape gets more deformed from a circular shape. The increasing value of parameter $\beta$ causes the size of the shadow to become smaller. To analyse the shadow, the required observables have been defined and plotted. One of the observables has been used to evaluate the energy emission rate for the rotating regular MOG black hole. From this, we have concluded that the peak of the energy emission rate shifts to a lower frequency for a relatively large value of the parameter $\beta$.
	
	It is hard to decouple the differential equations in terms of $r$ and $\theta$ without an equatorial approximation. We have reported the analysis of shadows using numerical techniques, using the observables as introduced by Hioki and Maeda. However, one can introduce new observables or can use other existing observables to study the shadows.
	
	In this work, we have demonstrated that classical regular black holes and regular horizonless dark compact objects, generally considered to be distinct families of astrophysical objects, are a family of connected astrophysical objects continuously deformed into one another, depending on the range and value of the
	parameter $\beta$. Our work illustrates that different strong gravity geometries describe alternative states of black holes and compact astrophysical objects in their lifetime. It is expected that at the small scale reached at the central value of the compact object when $r\rightarrow 0$, quantum gravity will take over\cite{Modesto:2010uh}.  The regular and horizonless compact objects derived from the MOG field equations in this work are classical in nature. The stability of photon orbits around the black hole and dark compact object shadows will produce viability issues for the existence of these astrophysical objects. For the MOG-Schwarzschild and MOG-Kerr solutions with two horizons, the inner Cauchy horizon can lead to instability problems.
	
	In future work, the gravitational collapse of stars will be investigated, assuming a form of matter and stress-energy, by solving the time dependent MOG field equations. Moreover, the merging of the regular and horizonless dark compact objects, producing gravitational waves and the subsequent ringdown phase, will be investigated. {Singularity-resolving physics in photon rings can further be studied in context of MOG theory\cite{Eichhorn_2023}.}
	
	{ 
		\section*{Acknowledgement}
	Research at the Perimeter Institute is supported by the Government of Canada through the Department of Innovation, Science and Economic Development Canada and by the Province of Ontario through the Ministry of Research, Innovation and Science.}
	\bibliography{References_2}

	\bibliographystyle{./utphys1}
\end{document}